\documentclass{desyproc}

\begin{document}
\title{Theory of Heavy-Ion Collisions}

\author{{\slshape Thorsten Renk$^1$}\\[1ex]
$^1$Department of Physics, P.O. Box 35, FI-40014 University of Jyv\"{a}skyl\"{a}, Finland}

\contribID{11}

\confID{1761}  
\desyproc{DESY-PROC-2009-xx}
\acronym{LP09} 
\doi  

\maketitle

\begin{abstract}
In high energy nucleus-nucleus (A-A) collisions, a transient state of thermalized, hot and dense matter governed by Quantum Chromodynamics (QCD) is produced. Properties of this state are reflected in the bulk low transverse momentum ($P_T$) hadron production which represent the remnant of the collective medium as well as in modifications of so-called probes which are not part of the thermalized medium, i.e. jets generated in high $P_T$ processes or leptons and photons which do not participate in the strong interaction. Theory effords aim at deducing the properties of QCD thermodynamics and collectivity from such observables.
\end{abstract}

\section{Introduction}

Often the aim of science is to understand the nature of a phenomenon in terms of its more fundamental constituents. This corresponds to a paradigm called 'reductionism', and the goal of high-energy physics can be understood from this paradigm as uncovering more fundamental building blocks of matter by probing ever decreasing distance scales. However, there are some phenomena in nature which require a different paradigm in order to understand them. Here, properties of a given system cannot be determined or explained by the sum of its fundamental constituents alone. Instead, the system as a whole determines in an important way how the parts behave. The corresponding paradigm has the name 'holism'.

An example in physics is Quantum Chromodynamics (QCD), the theory of strong interactions. While QCD at small distance scales is comparatively simple and can be understood using perturbation theory as the interaction of quarks and gluons as degrees of freedom, at large distance scales QCD shows phenomena like confinement and the appearance of hadrons as degrees of freedom which cannot easily be read off from the Lagrangean. Moreover, the thermodynamics of QCD matter appears quite complex in predicting a transition from a hadronic gas at low temperatures to a new state of matter with different degrees of freedom, the Quark-Gluon Plasma (QGP) above a transition temperature $T_C$ of about 170 MeV. The properties of this transition are accessible experimentally in ultrarelativistic heavy-ion collisions (URHIC) where matter with peak energy densities corresponding to temperatures above $T_C$ are reached for short times. Such experiments have been carried out at the CERN SPS at $\sqrt{s} = 17.6$ AGeV in the past, are currently being done at the Brookhaven National Lab RHIC collider at $\sqrt{s} = 200$ AGeV and will in the future be part of the CERN LHC program with Pb-Pb collisions at $\sqrt{s} = 5.5$ ATeV.

However, extracting these properties is not an easy task as always the system as a whole needs to be considered rather than an exclusive final state. Experimentally, this implies dealing with $O(10.000)$ particles in the detector while theoretically direct perturbative calculations from the QCD Lagrangean cannot be done --- even appropriate concepts to describe the system have to be found. 

It has proven useful to analyze A-A collisions in terms of 'bulk' and 'probes'. Here, 'bulk' stands for the part of the system which exhibits collectivity and is approximately thermalized. In terms of a momentum scale, this typically implies $P_T = O(\text{few}\;100 \,\text{MeV})$, i.e. of the order of $T_C$. Particles at much higher momentum scales never thermalize and hence cannot be treated as part of the bulk matter. However, they nevertheless interact with the medium, and hence can serve as probes of the medium. Typically, the presence of the bulk medium implies either production channels of probes or final state interactions which are not present in more elementary reactions like p-p collisions, hence the modification of probes as compared to the suitably scaled p-p baseline carries information about the medium. Examples for important probes in heavy-ion physics are high $P_T$ hadrons and jets, leptons and photons and heavy-quark bound states.

Using these concepts, one can examine heavy-ion collisions by looking at the bulk medium itself, by studying the modification of probes by the medium as compared to a p-p baseline, and finally also the modification of the bulk medium due to the interaction with a probe, i.e. its response to a local perturbation. 

\section{The bulk medium}

\begin{figure}[hb]
\centerline{\includegraphics[width=0.5\textwidth]{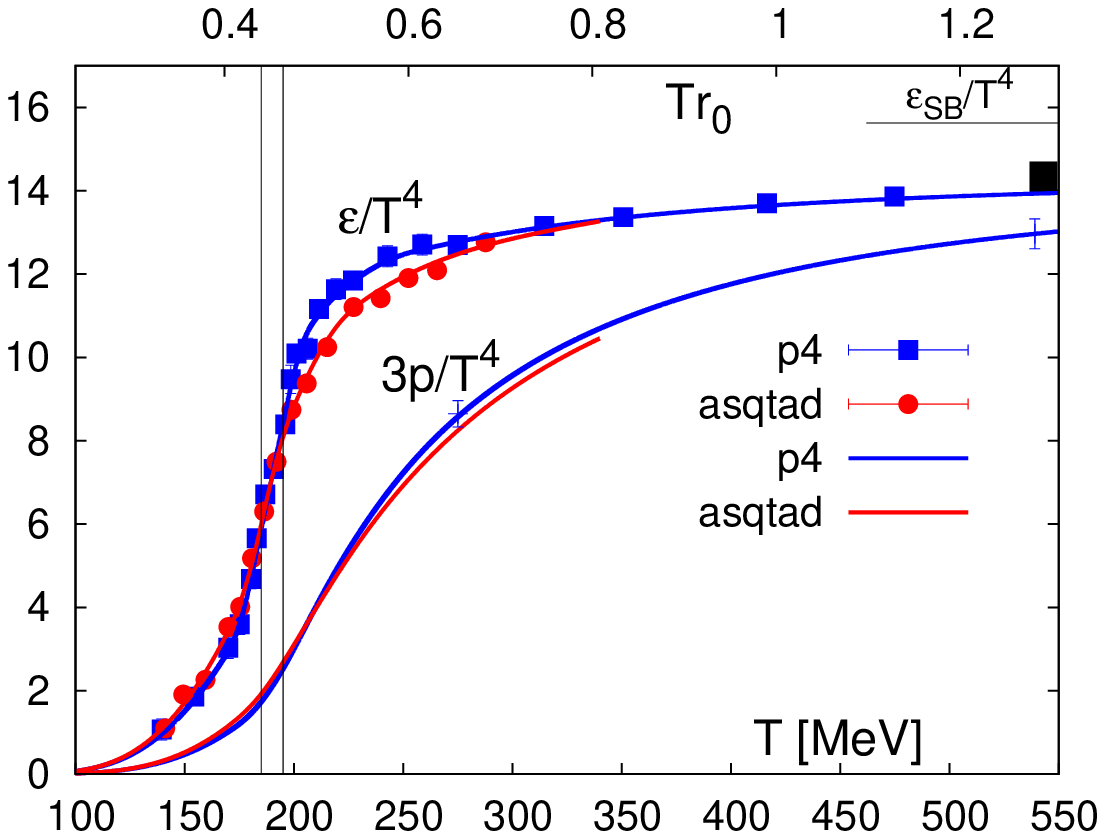}\includegraphics[width=0.5\textwidth]{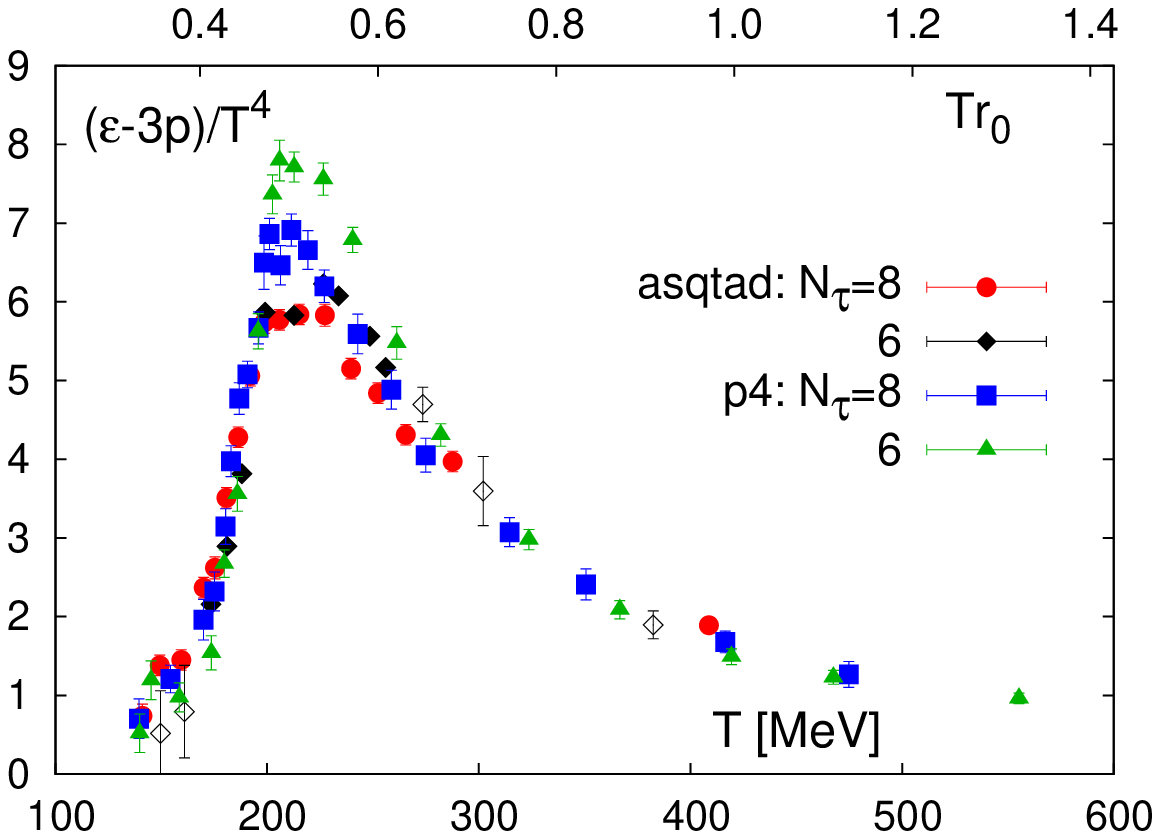}}
\caption{Lattice results \cite{Lattice} for equation of state (left) and interaction measure (right) of hot QCD.}\label{F-Lattice}
\end{figure}

Theoretical expectations about the thermodynamics of the bulk medium can be formed from lattice QCD simulations at finite temperature. While these can be done only for a static system, they allow to study thermodynamical properties of hot QCD. An example for such results \cite{Lattice} is shown in Fig.~\ref{F-Lattice} in terms of normalized energy density $\epsilon$ and pressure $p$ as a function of temperature $T$ and the so-called interaction measure $(\epsilon - 3p)$ which measures deviations from an ideal gas behaviour.

These results, in particular the strong change of thermodynamical properties around $T=170$ MeV, are indicative of a phase transition or a rapid crossover. The large value of the interaction measure indicates that the system is, at least close to $T_C$, far from an ideal gas and instead strongly coupled. However, the applicability of lattice results, which describe a static medium in termal equilibrium to the highly dynamical situation in heavy-ion collisions is not {\em a priori} obvious. What is observed in the detector are free hadrons long after the breakup of any bulk medium created in the collision. Any conclusions as to the formation of a medium hence need to rely on indirect evidence, i.e. on the imprint of such a medium on the final distribution of hadrons.

First evidence for collectivity is found in the mass ordering of transverse mass $m_T$ spectra (where $m_T = \sqrt{P_T^2+m^2}$ with $m$ the particle mass and $P_T$ its transverse momentum). In Pb-Pb collisions at the CERN SPS \cite{SPS} it was observed that transverse mass spectra of hadrons obey the formula

\begin{displaymath}
\frac{1}{m_T} \frac{dN}{d m_T} = \exp\left[-\frac{m_T}{T^*} \right] \quad \text{with} \quad T^*=T + m\langle v_T \rangle.
\end{displaymath}

Such mass ordering in which heavier hadrons are characterized by harder spectra is difficult to understand in terms of direct hadron production, but has a natural explanation in terms of collective motion of a thermalized volume with average collective velocity $\langle v_T \rangle$, thus the apparent temperature $T^*$ of the system has a part $T$ due to random motion and a part $m\langle v_T \rangle$ due to collective motion. This interpretation naturally leads to a fluid picture of the bulk medium in which individual fluid elements are locally thermalized and the fluid pressure drives the collective expansion of the system, till eventually the mean free path of hadrons inside the fluid becomes larger than the dimensions of the system, and decoupling into a system of free hadrons occurs. Hydro-inspired parametrizations (e.g. \cite{Blast,Fireball}) and later ideal relativistic hydrodynamical simulations (e.g. \cite{Hydro2d,Hydro3d}) have since been very successful at describing the various hadron spectra both in Pb-Pb collisions at the CERN SPS fixed target experiment at 17.3 AGeV and for Au-Au collisions at the Brookhaven RHIC collider at 200 AGeV.

Relativistic fluid dynamics is based on energy-momentum and current conservation, 

\begin{displaymath}
\partial_\mu T^{\mu \nu}  = 0 \qquad \partial_\mu j_i^\mu = 0 \quad \text{where} \quad T_{id}^{\mu \nu} = (\epsilon + p) u^\mu u^\nu - p g^{\mu \nu}
\end{displaymath}

with $j_i^\mu$ a conserved current (like e.g. baryon number), $u^\mu$ the 4-velocity of a fluid element, $\epsilon$ its energy density and $p$ its pressure, where properties of the medium enter in terms of the equation of state as e.g. the temperature dependence of the pressure $p(T)$. $T_{id}^{\mu \nu}$ here is the energy-momentum tensor of an ideal fluid with vanishing mean free path of the microscopic degrees of freedom. For finite mean free path, viscous corrections enter in the form $T^{\mu\nu} = T_{id}^{\mu\nu} + \Pi^{\mu \nu}$ where $\Pi^{\mu\nu}$ contains various gradients, e.g. a shear term which couples to velocity gradients. For a stable, causal result, gradients up to 2nd order have to be considered. In recent years, there has been tremendous numerical progress in the treatment of relativistic viscous hydrodynamics \cite{Viscous} going beyond the applicability of ideal hydrodynamics.

One of the most striking signatures of hydrodynamical behaviour is the so-called {\em elliptic flow}. If one makes a decomposition of the angular distribution of hadrons produced in a heavy-ion collision, elliptic flow appears as the second harmonic coefficient $v_2$,

\begin{displaymath}
\frac{dN}{d\phi} = \frac{1}{2 \pi} \left[ 1 + 2 v_1 \cos{\phi} + 2 v_2 \cos{2 \phi}+ \dots \right].
\end{displaymath}

In a hydrodynamical system, $v_2$ arises because the fluid pressure converts initial anisotropies in position space (such as present in the shape of the overlap region in non-central A-A collisions) to anisotropies in momentum space. The impact parameter dependence of $v_2$ is therefore a direct probe of pressure gradients in the system, and fluid dynamics must be able to give accurate predictions for $v_2$ as a function of impact parameter or $P_T$ if it is a valid description of the dynamics.

\begin{figure}[!hb]
\centerline{\includegraphics[width=0.5\textwidth]{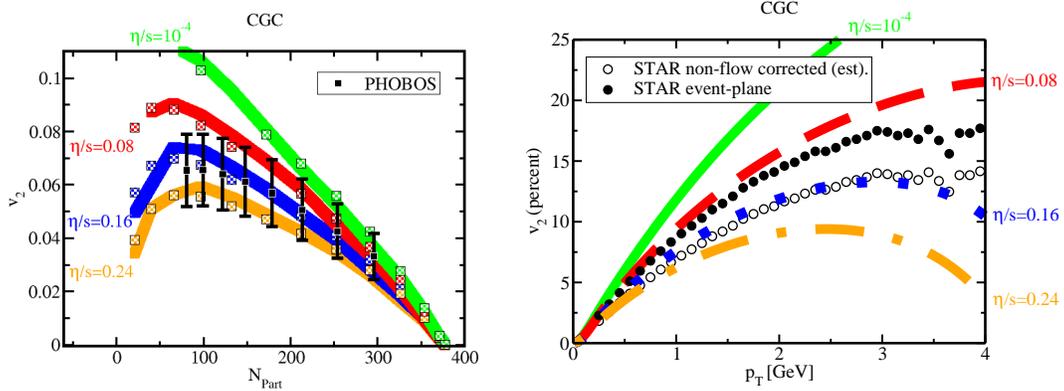}\raisebox{0.3cm}{\includegraphics[width=0.49\textwidth]{renk.v2-talk2.eps}}}
\caption{Relativistic viscous hydrodynamics results \cite{v2} for $v_2$ as a function of the number of collision participants (left) and $P_T$ (right) for different values of $\eta/s$ .}\label{F-Viscous}
\end{figure}

Viscous hydrodynamical results for $v_2$ have been obtained in recent years, and examples are shown in Fig.~\ref{F-Viscous}. The key parameter characterizing viscous effects is the ratio of shear viscosity $\eta$ over entropy density $s$.

From the figures, it is apparent that the data can be well described with a range of about $\eta/s = [0.08..0.2]$. This implies that the system is not ideal. However, one may note that superfluid helium is characterized by a value of $\eta/s$ about 10 times larger. This implies that the system created in heavy-ion collisions is very close to a perfect liquid, in fact it is the most perfect liquid known in nature, and that the mean free path in the medium is extremely small and the system hence exhibits strong collectivity. 

\section{Jet tomography}

The fluid picture applies for $P_T$ of $O(\text{few}\,T)$, however there are hadrons produced in heavy ion collisions with $P_T \gg T$. Such hadrons must come from hard, partonic processes. By arguments based on the uncertainty principle, one can establish that for typical kinematics the initial hard process takes place before a collective medium is produced and probes length scales at which collectivity is not relevant. Thus, the production of high $p_T$ partons is unmodified by the medium. However, the subsequent QCD evolution from a highly virtual initial parton into a parton shower at lower virtuality scales probes length- and timescales comparable with the medium lifetime and extent, therefore the parton shower is likely to be medium modified. Finally, the non-perturbative hadronization process can be safely estimated to take place outside the medium, therefore it is again unmodified by any final state interaction. This leads to a picture in which a hard probe with known and calculable properties is created in the medium and subsequently modified by its passage through the medium. The idea to exploit this effect in order to characterize medium properties is known as {\em jet tomography}.

Experimentally, the final state interaction of hard partons with the medium leads to very striking phenomena, among them the suppression of single inclusice high $P_T$ hadrons by a factor of about five in central 200 AGeV Au-Au collisions \cite{PHENIX-RAA} as compared to the scaled expectation from p-p collisions or the appearance of monojets in events in which one parton of a back-to-back event is absorbed by the medium \cite{STAR-Dijets}. The most commonly discussed observable is the nuclear suppression factor $R_{AA}$ of single high $P_T$ hadrons. It is defined as the hadron yield in A-A collisions divided by the yield in p-p collisions scaled with the number of binary collisions, \begin{displaymath}
R_{AA}(P_T,y) = \frac{d^2N^{AA}/dp_Tdy}{T_{AA}(0) d^2 \sigma^{NN}/dP_Tdy}.
\end{displaymath} 
In the absence of any nuclear initial of final state effects, $R_{AA}$ would hence be unity. Since the absence of strong initial state nuclear effects has been demonstrated in d-Au collisions, the strong deviation of $R_{AA}$ from unity can almost exclusively be attributed to final state interations of produced partons with the bulk medium. 

The likely mechanism for these modifications as compared to the p-p baseline expectation is medium-induced gluon radiation. The basic physics process is that gluons from the virtual gluon cloud surrounding a parton which propagates through the medium can decohere from the parent wave function if they pick up sufficient virtuality from the medium through interactions. A measure for the strength of the medium effect is the transport coefficient $\hat{q} = \frac{\langle q^2_\perp\rangle_{med}}{\lambda}$ which measures the average momentum broadening per unit pathlength $\lambda$. Since the phase $\phi$ of a gluon relative to the parent parton needs to be $O(1)$ for decoherence, one can estimate $\phi = \langle \frac{k_\perp^2}{2\omega} \Delta x\rangle = \frac{\hat{q}L}{2\omega}L = \frac{\omega_c}{\omega}$ where $\omega_c = \frac{1}{2} \hat{q} L^2$ is the characteristic scale of energy loss. This typical scale grows in a constant medium quadratically with pathlength. Based on similar estimates, the spectrum of radiated gluons per unit pathlength can be computed  to be $\omega \frac{dI}{d\omega dz} \sim \sqrt{\frac{\omega_c}{\omega}}$.

\begin{figure}[!hb]
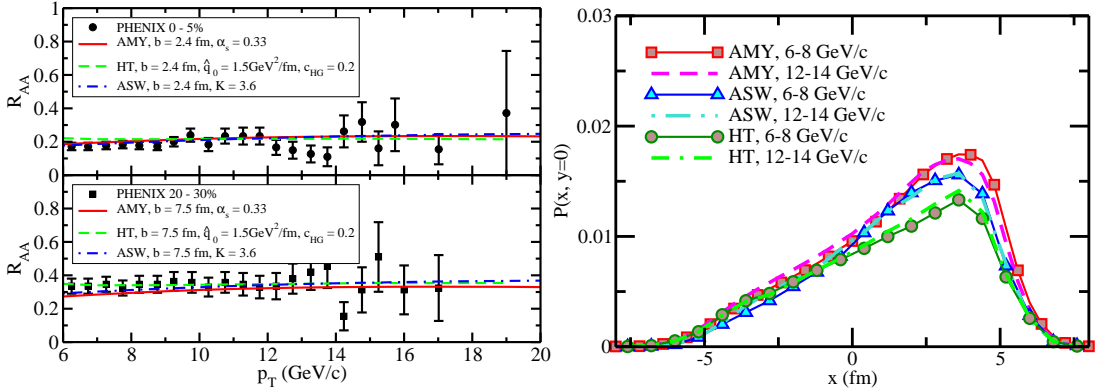

\centerline{\includegraphics[width=0.5\textwidth]{renk.RAA_centrality.eps}{\includegraphics[width=0.5\textwidth]{renk.RAA_Pxy0_b2.4.eps}}}
\caption{Nuclear suppression factor $R_{AA}$  in 200 AGeV Au-Au collisions (left) and distribution of vertex of origin inside the medium for observed high $P_T$ hadrons (right) for various models of radiative energy loss \cite{Eloss} in comparison with PHENIX data \cite{PHENIX-RAA}.}\label{F-RAA}
\end{figure}

As mentioned above, hard partonic processes typically lead to the creation of highly virtual partons which evolve into a parton shower. However, for measurements which focus on the observation of high $P_T$ hadrons there is a substantial bias towards events in which most of the energy within the shower is carried by a single hard parton. In this situation, it is justified to approximate the situation of a shower which develops inside the medium by a parton which loses energy into the medium while traversing it, i.e. subleading shower partons are not explicitly tracked in this approximation. Calculations utilizing this energy-loss picture in which hard parton trajectories are embedded into a full 3+1 dim fluid dynamical model for the bulk medium have reached a high degree of sophistication (see e.g. \cite{Eloss}). As apparent from Fig.~\ref{F-RAA}, they can reproduce both $P_T$ dependence and centrality dependence of $R_{AA}$ quite well and currently allow the extraction of information about the transport coefficient $\hat{q}$ from the medium with an accuracy of $\sim 50$\%.

The obvious next goal is to establish where the energy lost from the leading parton is recovered, and thus to confirm or disprove the picture of radiative energy loss. One possibility is a perturbative redistribution of energy within the parton shower --- energy lost from the leading shower parton would then lead to increased production of partons at low momenta. Currently, several Monte Carlo (MC) codes based on known vacuum shower codes like PYTHIA \cite{PYTHIA} or HERWIG \cite{HERWIG} are being developed \cite{Shower-MC}. They compute the whole medium-modified shower by including the possibility to have the parton kinematics or their branching probability modified by the interaction with the medium. Since one of the aims of jet tomography is to establish the relevant microscopic dynamics in the medium, currently the shower MC codes all include an assumption about the nature of the medium effect which eventually needs to be tested against data. However, there are consistent prescriptions to include generic quantum effects like the LPM suppression into the computation. Presumably, the LHC kinematic range is needed to observe clear jets and perturbative redistribution of energy.

\begin{figure}[hb]
\centerline{\includegraphics[width=0.5\textwidth]{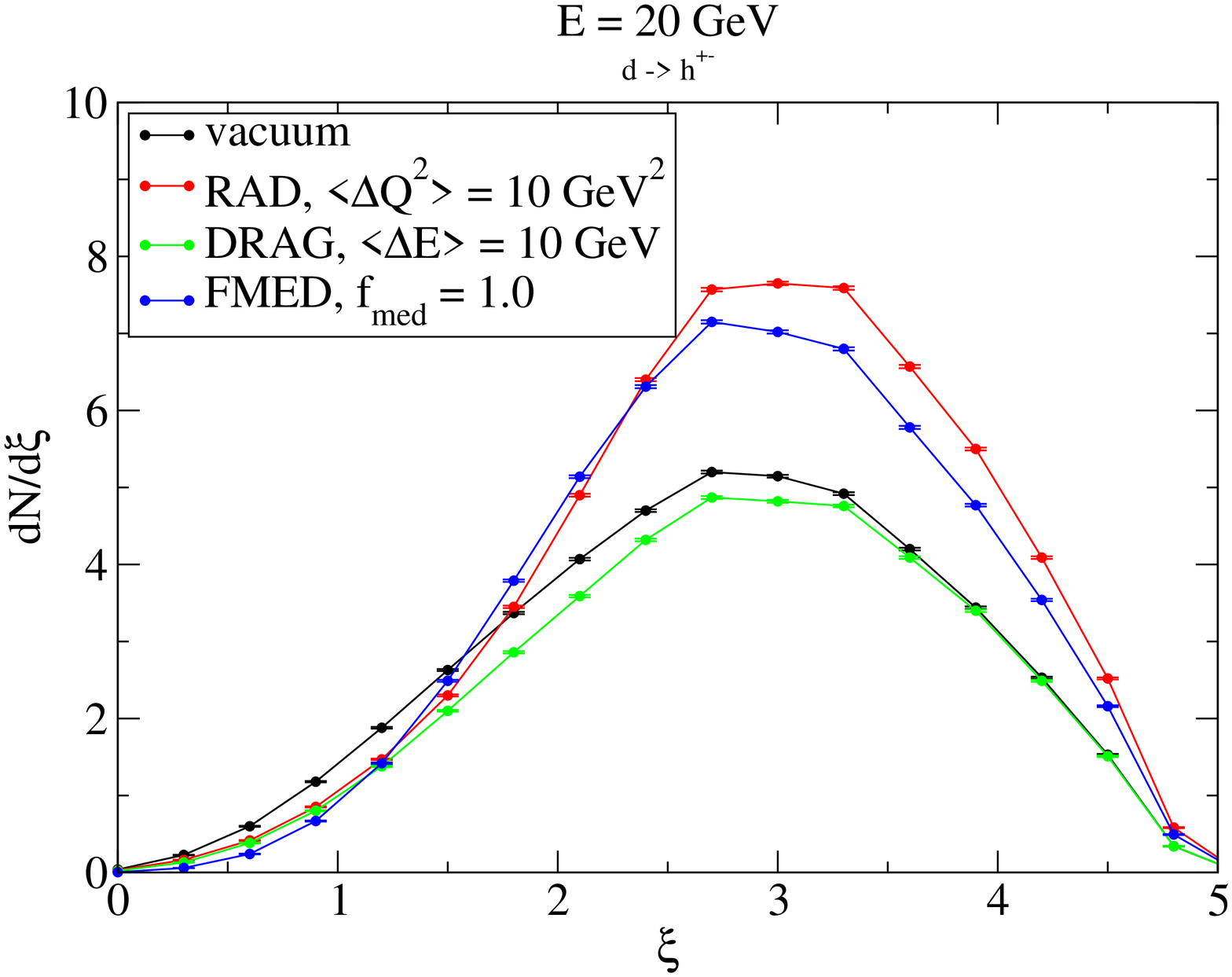}{\includegraphics[width=0.5\textwidth]{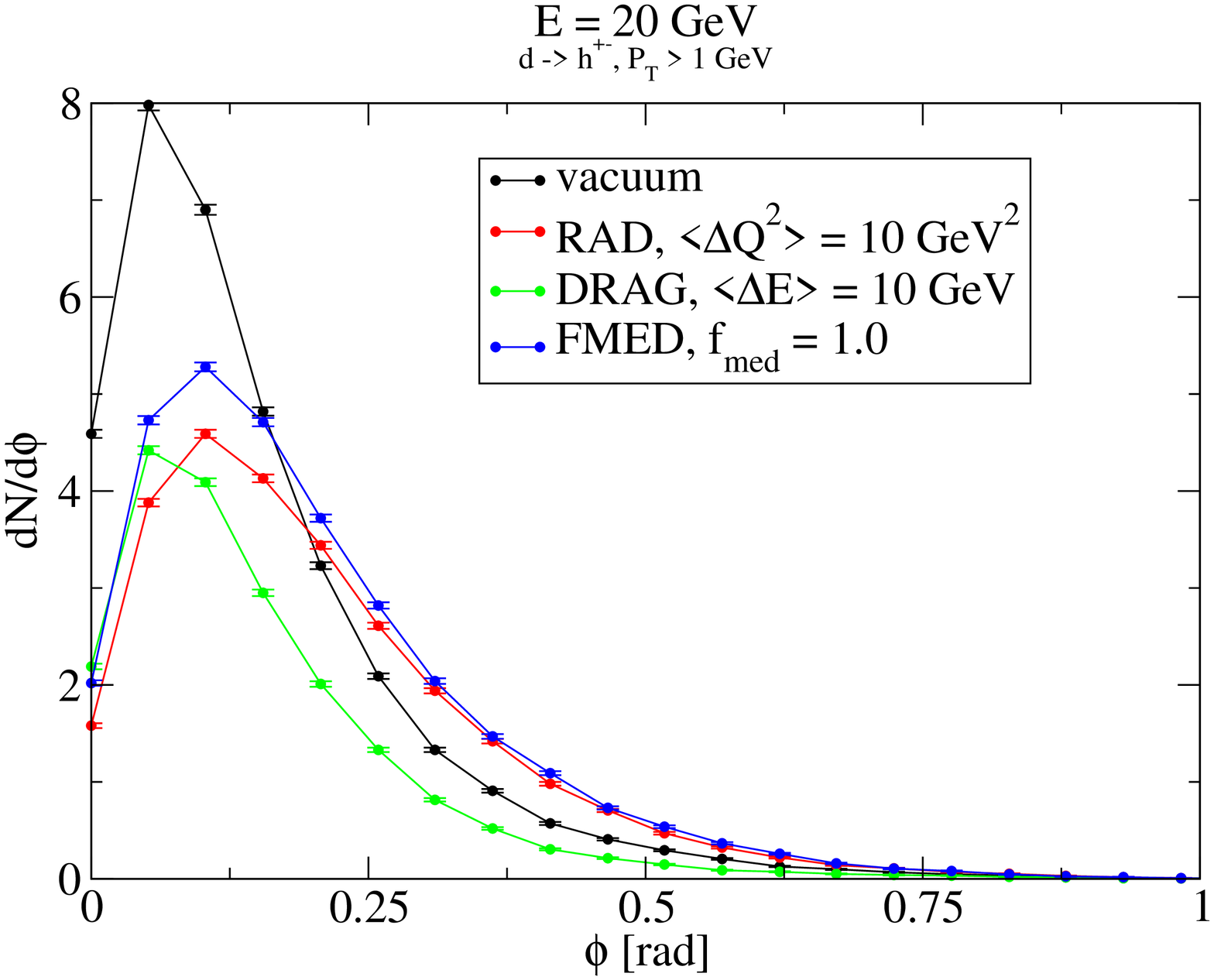}}}
\caption{Modification of the so-called hump-backed plateau $dN/d\xi$ where $\xi = \ln(1/x)$ and $x = p/E_{jet}$ (left) and of the angular distribution of hadrons in the jet (right) by various models for the jet-medium interaction \cite{Jetshapes}.}\label{F-Jets}
\end{figure}

Using such MC codes, the modification of typical jet observables like thrust, the subjet fraction, the jet shape or the longitudinal momentum distribution in the jet by the medium can be computed. Examples for the medium effect on such observables are shown in Fig.~\ref{F-Jets}. However, jet finding in the environment of a heavy-ion collision is very difficult due to the high level of background created by the bulk medium, therefore the bias introduced by the jet definition must be carefully studied before such medium effects can actually be observed and a comparison of the model results with data can be made.

Experimentally, at least the onset of the characteristic enhancement of low $P_T$ particle production induced by perturbative energy redistribution inside the jet cone should have been observable in $\gamma$-hadron correlations. The fact that this has not been seen so far points towards a different mechanism of energy redistribution being relevant at RHIC kinematics.

\section{Medium response}

\begin{figure}[hb]
\centerline{\includegraphics[width=\textwidth]{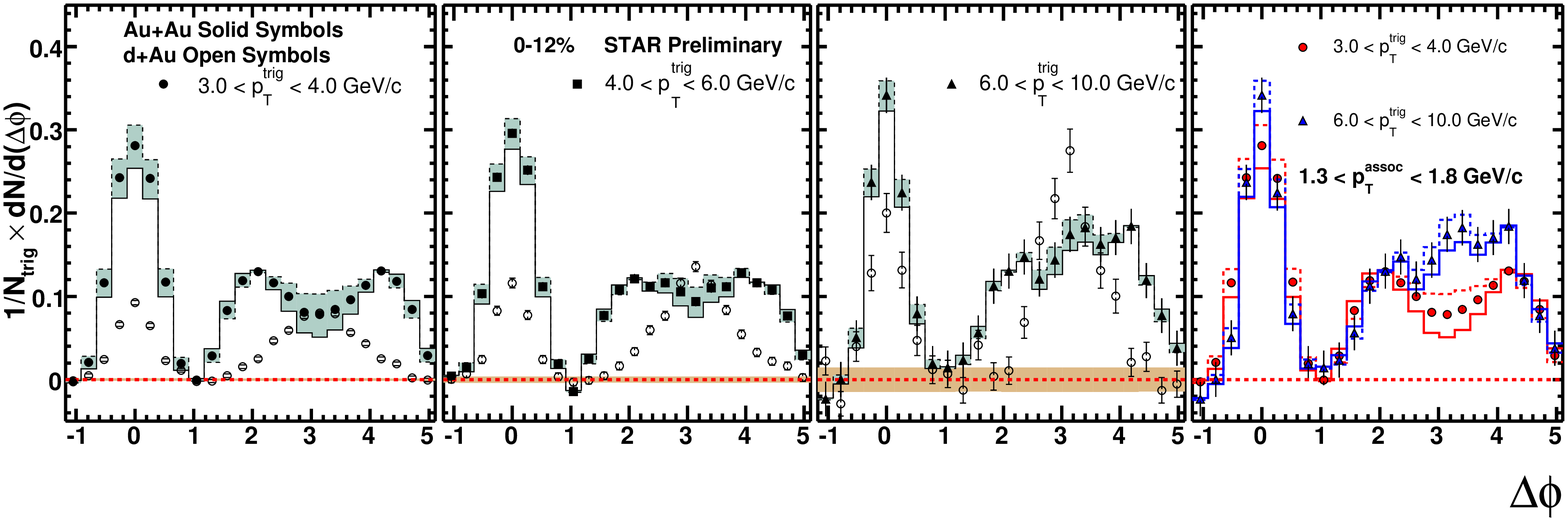}}
\caption{Correlations of particles with a high $P_T$ trigger hadron (defining $\Delta \phi=0$) as a function of angle for d-Au collisions (open symbols) and 200 AGeV Au-Au collisions (solid symbols) at RHIC \cite{STAR_cone} for rising values of associate hadron $P_T$.}\label{F-Cone}
\end{figure}

Measurements of the correlation strength of hadrons associated with a high $P_T$ trigger hadron have shed some light on a possible non-perturbative mechanism of energy redistribution. From these results (see Fig.~\ref{F-Cone}) it is apparent that without a medium (i.e. in d-Au collisions) the correlations reflect back-to-back jet events. However, especially at low $P_T$, the away side ($\Delta \phi=\pi$) correlation function in Au-Au collisions does not resemble a jet-like structure at all, rather it exhibits a characteristic double-hump structure, and only at significantly higher momenta is a jet-like correlation recovered.

These results have been widely interpreted as reflecting the recoil of the bulk medium from the hard probe. In this scenario, at least part of the energy lost from hard partons is contained in the medium in the form of a shockwave, where the characteristic cone structure of the shockwave leads to the double-hump structure in the angular correlation function. Note that if a fluid description of the medium is valid, shockwaves arise quite naturally from local perturbations of the medium.

Hydrodynamical calculations carried out under the assumption that the energy lost from a hard parton acts as a local source term of energy and momentum in the fluid dynamical equations have established that shockwaves leading to a characteristic double-hump structure in the correlation function can indeed be created (see e.g. \cite{Shocks}). However, at present these calculations are just a proof of concept --- in order to compare with the measured correlations, it is not sufficient to compute the energy deposition of a single parton. Rather, the bias for the production point of the trigger hadron in the medium must be determined from an energy loss calculation, based on this information energy deposition into the medium must then be computed, taking also into account the distortion of any shockwave by the collective expansion of the medium and the resulting bias for detecting the shockwave.

To date, no full hydrodynamical calculation has included these effects, but there is a phenomenological hydro-inspired model which has a proper averaging over the bias induced by energy loss and shockwave distortion by flow \cite{MyCones} which reproduces the two-particle correlations shown above.

A crucial test for the assumption that the observed signal is a shockwave is then a measurement of three particle correlations. Since the correlation measurement represents an average over many events, it is not evident that the double hump structure is created by the dynamics of a single event --- a situation in which a jet peak is displaced to one side in a single event could average to the same correlation. However, the three particle correlation signal is different. In particular, if the correlation function is plotted as a function of $\phi_1$ and $\phi_2$ where $\phi_i$ is the angle of a measured hadron with the trigger, for a displaced peak scenario correlation strength is only created on the diagonal $\phi_1 \approx \phi_2$, whereas a shockwave cone leads to characteristical off-diagonal structures.

\begin{figure}[hb]
\centerline{\includegraphics[width=0.72\textwidth]{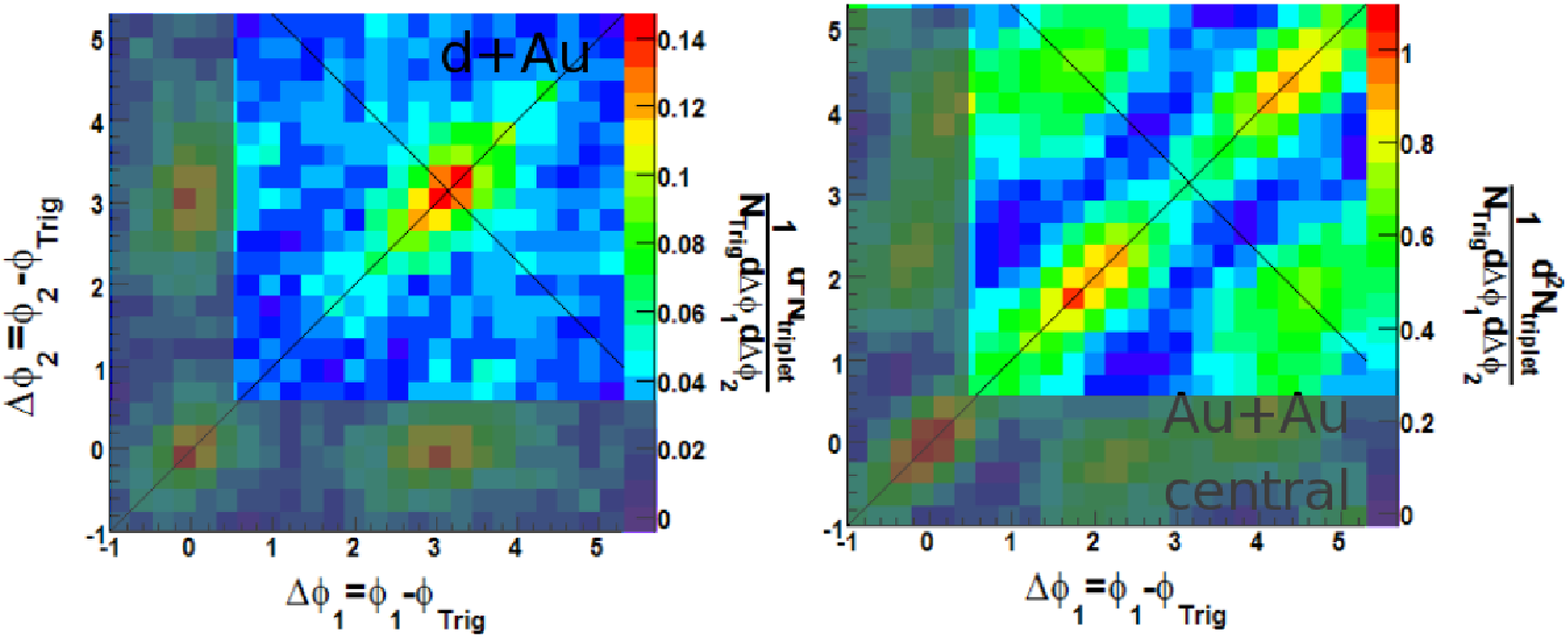}{\includegraphics[width=0.28\textwidth]{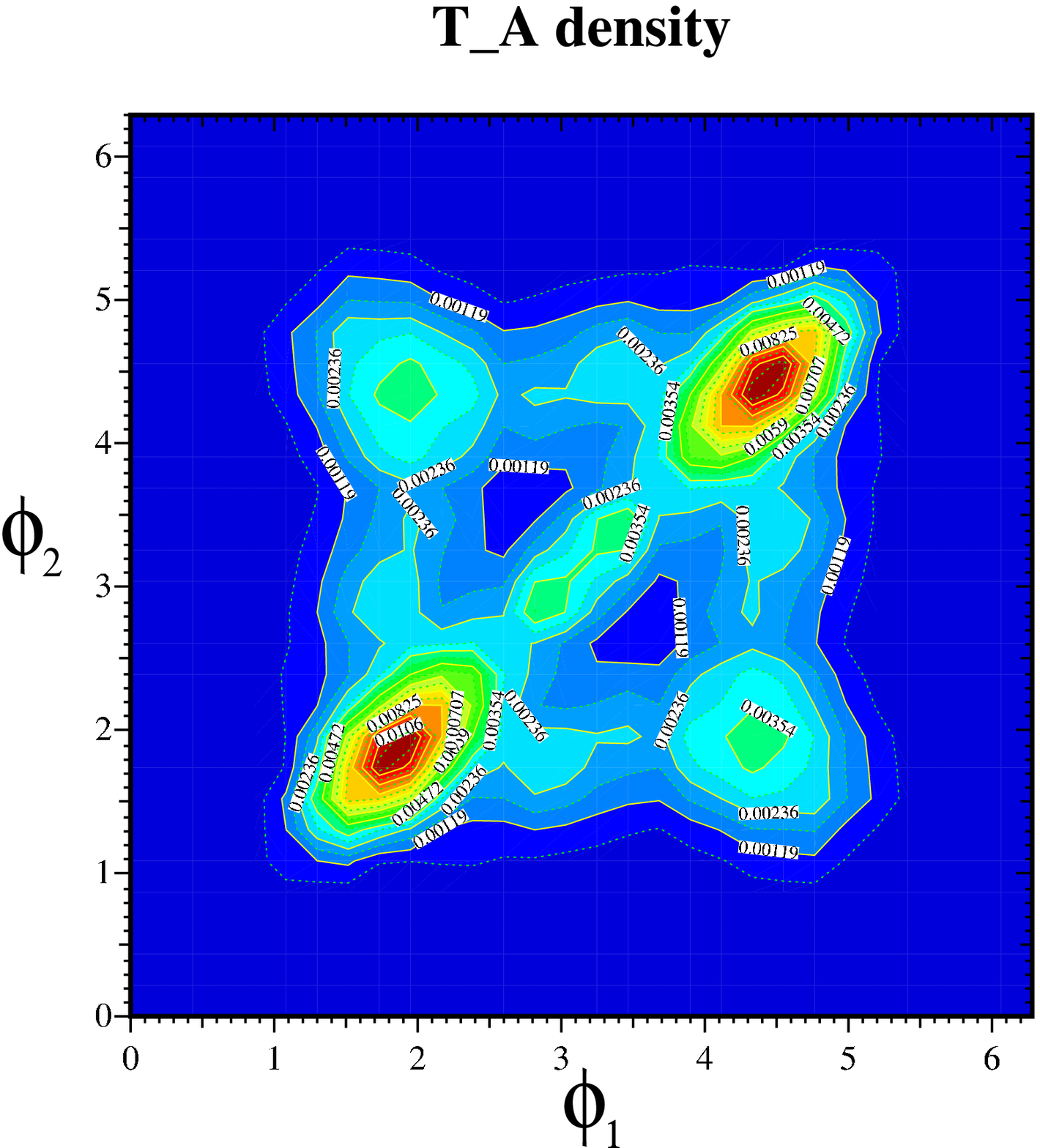}}}
\caption{Left: Correlations of two particles with a high $P_T$ trigger as measured in d-Au collisions (near side region is greyed out). Middle: Correlations of two particles with a high $P_T$ trigger as measured in central Au-Au collisions (near side region is greyed out) Right: Calculated correlation of two particles with a trigger in the away side region \cite{3pc}.}\label{F-3pc}
\end{figure}

In Fig.~\ref{F-3pc}, a comparison between three-particle correlations measured by the STAR collaboration in d-Au and Au-Au at 200 AGeV and a calculation for Au-Au \cite{3pc} for the away side is shown. While the d-Au measurement shows a signal very consistent with a back-to-back jet event, the correlation function in the Au-Au case exhibits a complicated structure on and off the diagonal. At least qualitatively the calculation manages to describe the observed signal well.

From these investigations can be inferred that the non-perturbative mechanism of shockwave excitation in the medium is a major channel absorbing energy lost from a leading parton. Conceptually, this is very interesting, as it offers in principle the possibility to measure various transport coefficients of the medium by observing its reaction to a localized perturbation. However, current theory efforts are still far away from this eventual goal.

Unlike at RHIC, at LHC energies multiple jet production per event is rather likely. Thus, shockwaves generated by the passage of hard partons through the bulk medium may actually become a major part of LHC bulk medium dynamics --- surprises are rather likely.

\section{Other major topics of interest}

Characterizing properties of the bulk medium via direct observation of bulk matter, jet tomography and medium recoil are however not the only interesting areas of research in theoretical heavy-ion physics. Another key questions is, that if the medium can be described as an almost perfect liquid, how it reached this state at all? In other words, by what interaction mechanism can a nuclear initial state equilibrate on a very short timescale? This question leads to topics like the description of the initial state in terms of low $x$ gluon saturation and the so-called Color-Glass-Condensate (CGC), which is believed to be the relevant state in the nuclear wave function. The subsequent 'shattering' of the CGC in the collision process next leads to a system with a very anisotropic distribution of particles in momentum space, which needs to isotropize before it can thermalize. Here, the physics of anisotropic coloured plasmas and plasma instabilities are research goals, with the ultimate aim to understand the onset of collectivity and to compute the initial state for hydrodynamical models.

A different set of questions is centered around the restoration of chiral symmetry, which in lattice QCD simulations takes place at the same temperature as the deconfinement transition from a hadron gas to a QGP. The chiral restoration requires that the vector correlator becomes degenerate with the axial correlator, but in what way this takes place is an open question. Experimentally, the vector mesons $\rho, \omega$ and $\phi$ as resonances in the vector correlator are most easily accessible, and their electromagnetic decays into dileptons offer the possibility to study their in-medium modifications. Such modifications may involve a shift in pole mass, as suggested by the so-called Brown-Rho scaling scenario, as well as a broadening of the meson widths. Theoretical studies of dilepton production within a fluid-dynamical model for the bulk medium in comparison with data suggest that chiral restoration is realized by dissolving resonance structures into a flat, featureless continuum. However, such calculations are rather involved and there is no clear consensus as to details yet.

Finally, in recent years the AdS/CFT correspondence discovered from String Theory has provided a new tool to compute properties of particular gauge theories in the strong coupling limit. In QCD, the strong coupling limit is very hard to access, thus the arrival of such methods has generated a lot of excitement. However, it remains to be seen how closely the gauge theories tractable by AdS/CFT methods resemble QCD, as present calculations are done for an $N=4$ SYM theory which does not exhibit running coupling, a chiral transition or a deconfinement transition, i.e. which omits almost all the interesting features of QCD whose study is the aim of heavy-ion physics in the first place.  

A good overview over topics currently relevant for heavy-ion physics can be found in the program of the Quark Matter 2009 conference \cite{QM}.

\section{Collectivity in QCD and the LHC}

If the aim of ultrarelativistic heavy-ion physics is the study of collectivity in QCD, and hence phenomena which take place mainly at a momentum scale of $O(T_C) \sim 0.2$ GeV, one may ask why this needs to be studied at the LHC which will provide collisions between lead ions at 5.5 ATeV, i.e. at a momentum scale several orders of magnitude above the scale at which collective phenomena take place. 

Part of the answer to the question is apparent from what has been said above: Techniques like jet tomography rely on the presence of hard processes in an event, and the abundance of high $P_T$ probes increases significantly with increased $\sqrt{s}$. However not only the quantity of hard probes increases, but also their quality: While reliable jet finding and the characterization of jet properties is difficult in a kinematic region where the jet energy is $O(20)$ GeV while the background is $O(1)$ GeV, this is no longer the case at LHC kinematics where jets with energies $O(500)$ GeV can be observed above a background with momentum scales $O(2-3)$ GeV. In addition, the kinematic range of the LHC offers access to processes like $Z^0$-jet back-to-back events, which are very clean probes as the narrow $Z^0$ decay signal can be detected practically background-free. Such probes allow a complete characterization of the jet kinematics independent from jet finding in the background, and hence can be used for precision calibration of the models.  

\begin{figure}[hb]
\centerline{\includegraphics[width=0.5\textwidth]{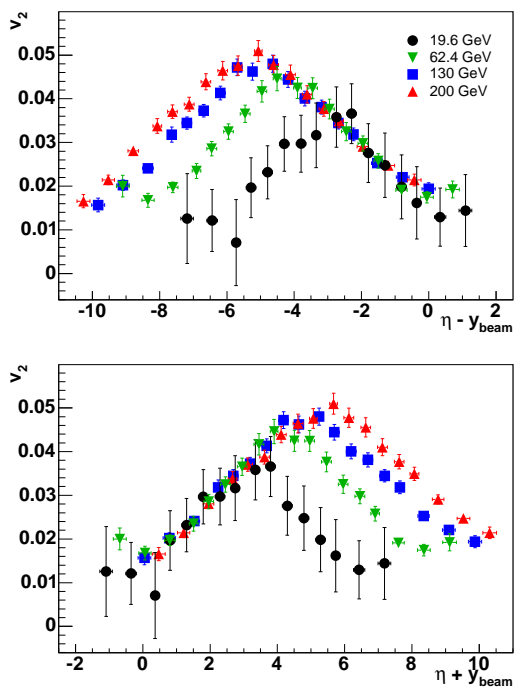}}
\caption{Elliptic flow coefficient $v_2$ as a function of rapidity difference to the beam rapidity for various values of $\sqrt{s}$ as obtained by the PHOBOS collaboration \cite{Phobos}.}\label{F-v2eta}
\end{figure}

However, jet tomography is not the only reason that collider kinematics is useful to probe collectivity in QCD --- the excitation function of the bulk medium itself. An example is shown in Fig.~\ref{F-v2eta}.

Here, the elliptic flow coefficient $v_2$ is shown as a function of the rapidity difference with the beam rapidity for various values of the collision energy $\sqrt{s}$. It is evident that $v_2(\eta\pm y_{beam})$ exhibits a characteristic triangular shape and a striking scaling behaviour. The change in this quantity is a very slow function of $\sqrt{s}$ --- while the excitation function of hard probes is given by $\sqrt{s}/2$, collective phenomena typically scale like $\log{\sqrt{s}}$, i.e. one needs a large kinematic lever-arm to observe the excitation function of collective phenomena at all.

For the particular observable $v_2(\eta-y_{beam})$, the question is if the scaling persists at LHC energies. The hydrodynamical picture predicts that the scaling will {\em not} be observed --- yet the scaling behaviour in itself looks simple and compelling. The (dis-)agreement of the measured excitation function of this observable will therefore play a crucial part in either confirming or disproving ideas about the dynamical picture underlying heavy-ion collisions. However, to make such an argument, the large extension in kinematical range provided by LHC is absolutely crucial. 

\section{Outlook}

What can be expected from future heavy-ion physics at the LHC? First of all, the huge extension in the kinematic range will help in our understanding of both bulk phenomena and jet tomography. Especially the physics of the interaction of hard probes with the medium will benefit enormously from the abundant production of high $P_T$ particles and from the access to very clean channels.

In a broader sense, while qualitatively the dynamics of a collective QCD medium can be understood in terms of the near-perfect liquid, quantitatively many features of the dynamical evolution are not yet well understood, and in many areas even qualitative tests of our understanding of the relevant physics mechanisms are needed. LHC results will, from the first day on, have a large impact on heavy-ion theory in terms of ruling out or confirming existing ideas. 

Clearly, there may be some surprises, for example there are hints that the dynamics of bulk recoil from hard probe energy loss may be an important feature of bulk dynamics at LHC, something that is not appreciated in predictions so far. But finally, there may be also genuinely new phenomena of collectivity in QCD to be discovered. 

\section*{Acknowledgements}

I'd like to thank Kari Eskola and Urs Wiedemann for their role in the preparation of this presentation. 
This work was supported by an Academy Research Fellowship from the
Finnish Academy (Project 130472) and from Academy Project 115262.
 

\begin{footnotesize}



%

\end{footnotesize}


\end{document}